# Growth and physical property study of single nanowire (diameter ~ 45nm) of half doped Manganite

Subarna Datta[1], Sayan Chandra[2], Sudeshna Samanta[1], K. Das[1], H. Srikanth[2], Barnali Ghosh[1*]

[1]Unit for Nanoscience, Department of Condensed Matter Physics and Material Sciences,
S.N. Bose National Centre for Basic Sciences, Salt Lake, Kolkata-700098, INDIA.

[2]Department of Physics, University of South Florida, Tampa, Florida 33620, USA.

Email: barnadutt@gmail.com
* Corresponding author: barnali@bose.res.in

**Abstract:**

We report here the growth and characterization of functional oxide nanowire of hole doped manganite of $La_{0.5}Sr_{0.5}MnO_3$ (LSMO). We also report four-probe electrical resistance measurement of a single nanowire of LSMO (diameter ~ 45nm) using Focused ion beam (FIB) fabricated electrodes. The wires are fabricated by hydrothermal method using autoclave at a temperature of 270 °C. The elemental analysis and physical property like electrical resistivity are studied at individual nanowire level. The quantitative determination of Mn valency and elemental mapping of constituent elements are done by using Electron Energy Loss Spectroscopy (EELS) in the Transmission Electron Microscopy (TEM) mode. We address the important issue of whether as a result of size reduction the nanowires can retain the desired composition, structure and physical properties. The nanowires used are found to have a ferromagnetic transition ($T_C$) at around 325 K which is very close to the bulk value of around 330 K found in single crystal of the same composition. It is confirmed that the functional behavior is likely to be retained even after size reduction of the nanowires to a diameter of 45 nm. The electrical resistivity shows insulating behavior within the measured temperature range which is similar to the bulk system.

## 1. Introduction

Nanowires of functional oxide materials because of their unique one-dimensional like structural characteristics and size effects exhibit many novel physical properties that are different from their bulk counterparts. It had been shown before that down to size range of 40 nm; the ferromagnetic $T_C$ is enhanced in nanowires of $La_{0.67}Ca_{0.33}MnO_3$ [1]. In past studies nanowires of $La_{0.67}Ca_{0.33}MnO_3$, were grown within templates [1, 2]. The enhancement in $T_C$ of nanowires corroborates the enhancement in $T_C$ of nanoparticles of same materials [3]. Our motivation is to investigate in what extent nanowires of complex oxides such as manganites, retain their functionality and physical properties on size reduction when the wires are grown by hydrothermal method. The 1D nano structures with well-controlled size, phase purity, crystallinity and chemical composition are synthesized by hydrothermal method in this paper. As a result, the retention of the composition, structural and physical properties on size reduction are important issues that need to be established. Often such structural characterizations are done at average level on an ensemble of nanowires. However, characterizations at the level of a single nanowire using spatially resolved tools are needed. In the present work we have used spatially resolved technique like TEM based EELS to investigate the chemical composition along with other structural and microscopic tools. Importantly we are able to carry out four-probe electrical measurement using FIB grown contacts on a single nanowire of diameter ~ 45nm.

The nanowires of functional oxides can be fabricated by different methods like template- assisted growth along with chemical solution processing [4], laser-assisted vapor-liquid-solid phase growth [5], lithography, solvothermal method [6] and hydrothermal process [7] etc. All the above mentioned methods have their own advantages and disadvantages. We have already reported template assisted growth of functional oxide nanowires by sol-gel synthesis method and magnetic properties of such nanowires [3, 8]**.** We report here template free fabrication of single-crystalline nanowires (~ 50 nm, length ~ 1-10 μm) of hole doped manganite of $La_{0.5}Sr_{0.5}MnO_3$ (LSMO) using hydrothermal method. Though this method is suffering from size (diameter) dispersion, it has the advantage of high through put.

## 2. Experimental Details:

Functional oxide nanowires of hole doped manganite $La_{0.5}Sr_{0.5}MnO_3$ (LSMO) was fabricated by hydrothermal synthesis using autoclave. $MnCl_2, 4H_2O$, $La(NO_3)_3, 6H_2O$ and $Sr(NO_3)_2$ were used as precursor materials, KOH was served as mineralizer, while $KMnO_4$ was used as oxidizer. The precursor materials were dissolved in de-ionized water, KOH was added while stirring to adjust the alkalinity of the solution [7]. The initial mole ratios of the input species were 0.6 $KMnO_4$:1.4 $MnCl_2.4H_2O$:1.0 $La(NO_3)_3 \cdot 6H_2O$:1.0 $Sr(NO_3)_2$:400 $H_2O$:70 KOH. The reaction mixture was vigorously stirred and poured into an autoclave. The main reaction was done in a Teflon vessel which has to be placed in a stainless steel autoclave for higher temperature hydrothermal treatment. The Teflon vessel was filled till 80% of its volume, and then this vessel was placed in the stainless steel container. The crystallization reaction was performed at 270 $^oC$ for 30 hours. After the reaction, the autoclave was cooled and depressurized; the products were washed with de-ionized water and dried in an oven at 120 °C for overnight. A black powder that contains the nanowires was finally obtained.

Nanowires were characterized by using different characterization tools; like, X-ray diffraction, Scanning Electron Microscope (SEM), Energy Dispersive analysis of X-ray (EDAX), High Resolution Transmission Electron Microscopy (HRTEM). The Phase formation and phase purity were confirmed by powder diffraction X-ray. The data was taken by PANalytical X'Pert PRO Laboratory powder X-ray diffractometer with Cu Kα radiation. The diameter and length of the wires are around 20-50 nm and 1-10 micron respectively, confirmed by Quanta FEI 200 Scanning Electron Microscope (SEM) and 200KV Tecnai G2 TF-20 Transmission Electron Microscope. The stoichiometry of the compound LSMO was determined using TEM-EDAX.

To establish the retention of composition on size reduction, we have used TEM based Electron Energy Loss Spectroscopy (EELS) to investigate the chemical composition at a single nanowire level [10]. 200 kV Tecnai G2-TF-20 TEM with a Gatan parallel detection EELS spectrometer was used for this study. The unknown Mn valency of LSMO nanowire is estimated from EELS. Energy Filtered Transmission Electron Microscopy (EFTEM) mode was done to image the distribution of constituent elements within the nanowire. In addition to the basic characterizations the magnetic nature of the nanowires was determined by a commercial Physical Property Measurement System (PPMS).

To make the four-wire electrical contacts on a single nanowire, we have dispersed the nanowires on $SiO_2$ substrate. A single nanowire of LSMO of diameter ~ 45 nm was connected to Cr/Au contact pads by interconnectors made of Pt deposited by FIB (FEI -HELIOS 600) using Ga ions at a voltage of 30 keV and beam current of 80 pA. The temperature variation of resistivity of a single nanowire down to liquid helium temperature (5 K) was done by using cryogenic cryocooler model SRDK 305.

**3. Results and discussions:**

*3.1. Structural and elemental characterization:*

A typical collection of nanowires obtained from autoclave is given in Fig. 1. Phase formation and phase purity were checked with X-Ray diffraction measurement shown in Fig. 2. It is the tetragonal structure of the space group *I*4/*mcm* and it compares well with bulk values [9] and it matches well with the data reported in ICSD pattern, ref code: 01-089-0786. The lattice parameters are a = 0.545 nm, b = 0.545 nm and c = 0.770 nm. From the structural characterization (shown in Fig. 2) of hydrothermally grown nanowires, it has been observed that we could form single tetragonal phase of LSMO (x = 0.5) nanowires which has no impurity.

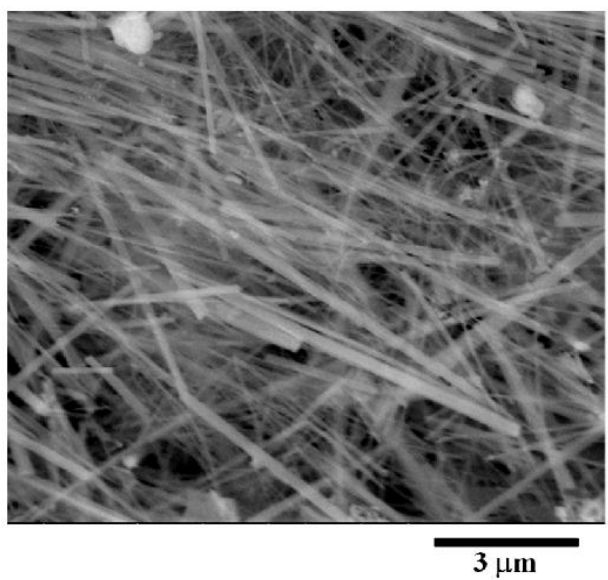

Fig. 1. SEM image of collection of nanowires

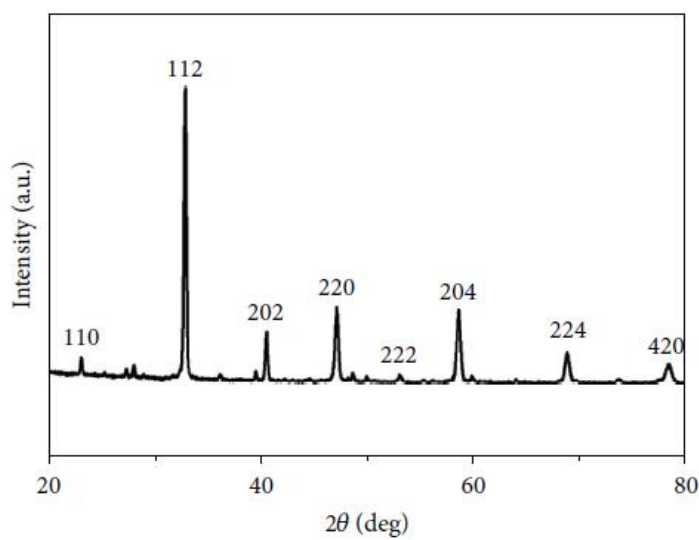

Fig. 2. XRD pattern of $La_{0.5}Sr_{0.5}MnO_3$ nanowires confirms the phase formation and phase purity.

*3.2. Transmission Electron Microscopy:*

The wires were grown of different diameters and lengths ranges from 20-50 nm and 1-10 micron respectively. The TEM image of a single nanowire is shown in Fig. 3(a). The single crystalline nature of the nanowires was confirmed from the diffraction pattern and HRTEM images shown in Fig. 3(b) and (c). The lattice spacing is around 3.11Å and (hkl) values are estimated from diffraction pattern data. Nanowires fabricated are of pure phase and single crystalline in nature. It has been observed from the TEM-EDAX data that the atomic percentage ratio (La:Sr):Mn:O is approximately 1:1:3 which is the desired composition. We have done the TEM analysis on many single nanowires across the length of the nanowires and it shows single crystallinity throughout the length.

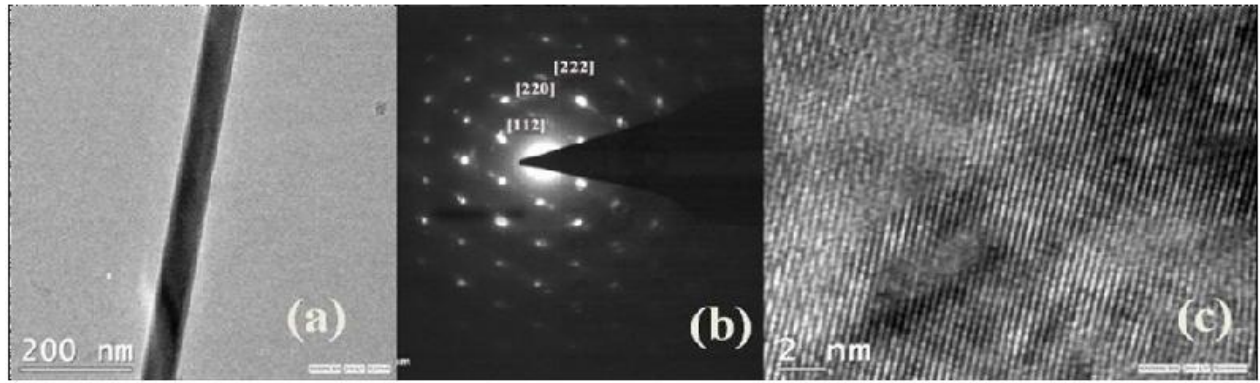

Fig. 3. (a) Single nanowire of diameter around 45 nm, (b) the selected area diffraction pattern taken with a Transmission Electron Microscope and (c) HRTEM image shows the single crystalline nature of LSMO nanowires.

*3.3. Elemental analysis using EELS:*

The elemental analysis of these nanowires was done by EELS on different single nanowires repeatedly and estimated the valency of Mn from the calibration curve shown in Fig. 4(a). We have determined Mn valency using the white lines ($L_2$, $L_3$ ionization edges of Mn) and the intensity ratio of $L_3$ and $L_2$ lines [11]. The intensities of $L_3$ and $L_2$ lines are related to the unoccupied states in the 3d bands. Transition from Mn 2p shell is actually split into two components separated by spin orbit splitting of the ionized 2p core level. Transition from $2p^{3/2}$ to $3d^{3/2}$ $3d^{5/2}$ and from $2p^{1/2}$ to $3d^{1/2}$ are $L_3$ and $L_2$ lines respectively. Comparing the intensity $L_3/L_2$ ratio of Mn of LSMO nanowire and that of Mn with known valency of some compounds, quantitative determination of Mn valency of our sample was evaluated [10]. The intensity ratios of $L_3$ and $L_2$ lines of different Mn oxide compounds as a function of their known valency are plotted in Fig. 4(a). This curve serves as the calibration curve from which the valence state of unknown materials can be obtained by using the observed intensity ratios. From the calibration curve, we have estimated the valency of Mn of LSMO nanowire. The Mn valency is of the order of ~ 3.5 and it is very close to its bulk value as shown in the Table 1. Energy Filtered TEM image was taken to check the homogeneity of the elemental distribution in each nanowire. Fig. 4(b) shows the EFTEM image of LSMO nanowire, where red, green, blue and yellow colors are used for elements O, Mn, La and Sr respectively. The EFTEM analysis shows that all the constituent elements La, Sr, Mn, and O are homogeneously distributed within the nanowire.

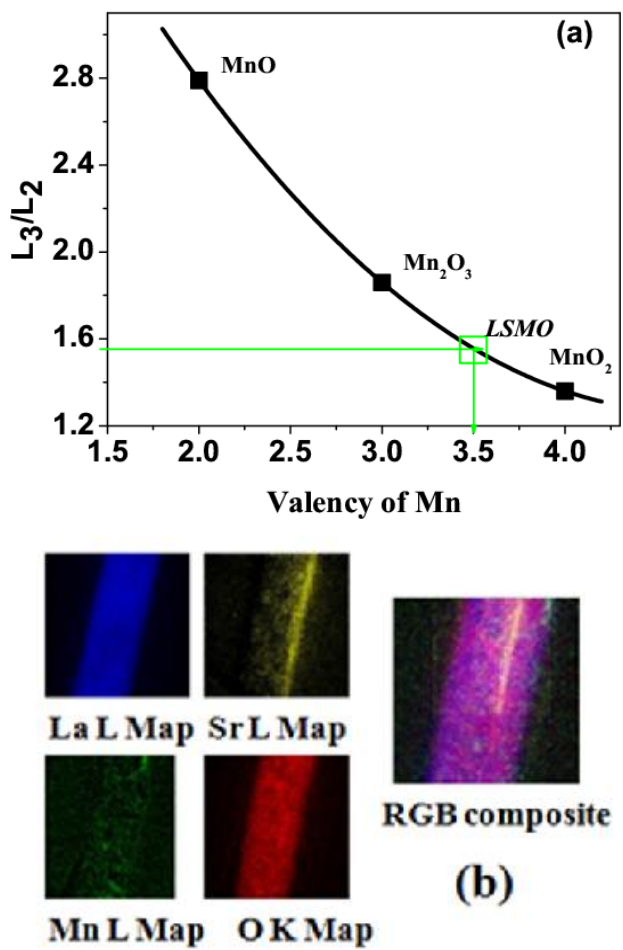

Fig. 4. (a) Intensity ratios of $L_3$ and $L_2$ lines of different Mn oxide compounds as a function of their known valency in the energy range 600-1000eV and (b) Energy Filtered Transmission Electron Microscope (EFTEM) image of each constituent element: L map of La, Sr, Mn and K map of O in LSMO nanowire. The red, green, blue and yellow colors represent the distribution of elements O, Mn, La and Sr respectively in the nanowire in the RGB composite map.

Table 1. Comparison of $T_C$ and valency of LSMO nanowires with the bulk.

| Sample | $T_C$ (K) | Mn valency |
|---|---|---|
| LSMO bulk | 330 | 3.5 |
| LSMO nanowire | 325 | 3.5 |

*3.4. Magnetic measurements:*

The results of magnetic measurements are shown in Fig. 5. It has been observed from the temperature variation of magnetization data that the LSMO nanowires undergo a transition from ferromagnetic (FM) to paramagnetic (PM) phase around 325 K shown in Fig. 5(a). The observed $T_C$ was compared with the phase diagram of bulk LSMO [12], shown in Table 1. Field variations of magnetization (M vs. H) were done at 300 K and 5 K in Fig. 5(b). The nanowires show ferromagnetic behavior with large coercivity ~ 645 Oe at 5 K shown in the inset of Fig. 5(b).

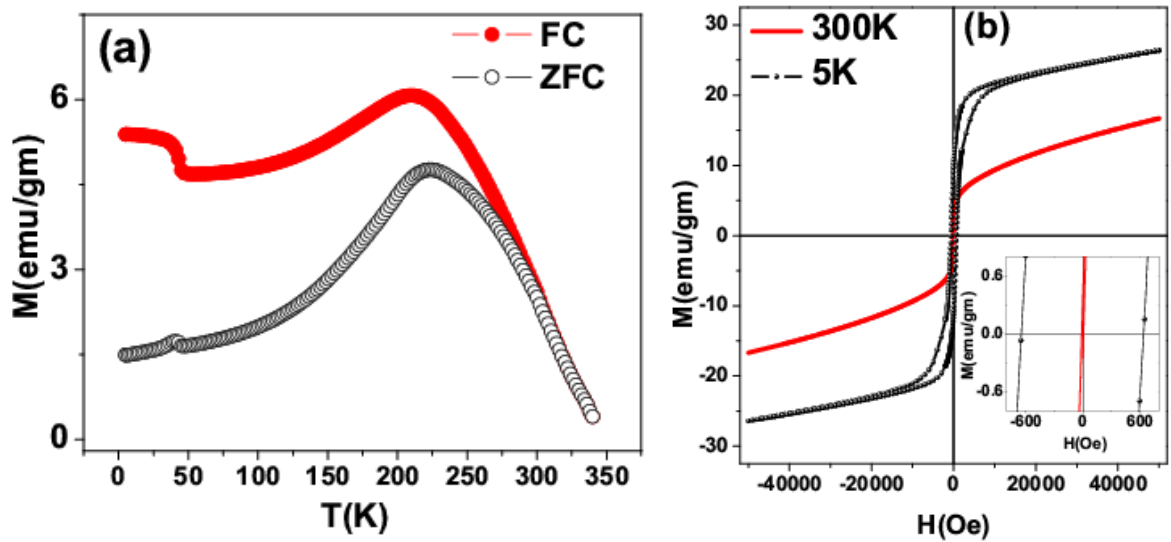

Fig. 5. (a) Temperature variation of magnetization data of LSMO at 100 Oe and (b) Plot of field variation Magnetization of LSMO at 5 K and 300 K.

Based on the magnetic data and the work on magnetocaloric study on these LSMO (x = 0.5) nanowires [13], we have observed the signatures of three distinct magnetic transitions from the change in magnetic entropy $-\Delta S_M$ (T): two maxima at 290 K and 45 K; minimum at 175 K. The maximum at 290 K arises due to PM-FM transition and around 175 K the transition is due to antiferromagnetic (AFM) transition. The above two transitions are also observed in bulk LSMO (x = 0.5) sample [14-16]. Around 45 K there is one maximum in the magnetization curve, this behavior is not seen in bulk. We have explained this transition as FM transition as observed in the magnetic entropy change plot [13]. From the magnetic measurements and from our previous magnetocaloric study [13]; we can infer that the hydrothermally grown nanowires show the basic ferromagnetic property as seen in bulk.

*3.5. Electrical resistivity measurement:*

*3.5.1 Fabrication of contact pads using FIB:*

For resistivity measurement Cr/Au contact pads are deposited on $SiO_2$ substrate by thermal evaporation using hard mask as shown in Fig. 6(a). Using Helios Dual beam system consisting of a FEG source and an ion beam source, we have located one single nanowire of diameter ~ 45nm as shown in Fig. 6(b). The electrical contact pads were deposited by using a Pt source by Focused Electron Beam (FEB) and then these four pads were finally connected to the pre-fabricated gold contact pads by FIB, shown in Fig. 6(c). The separation between two consecutive probes is around ~ 300 nm. The each contact pad has width of ~ 300 nm.

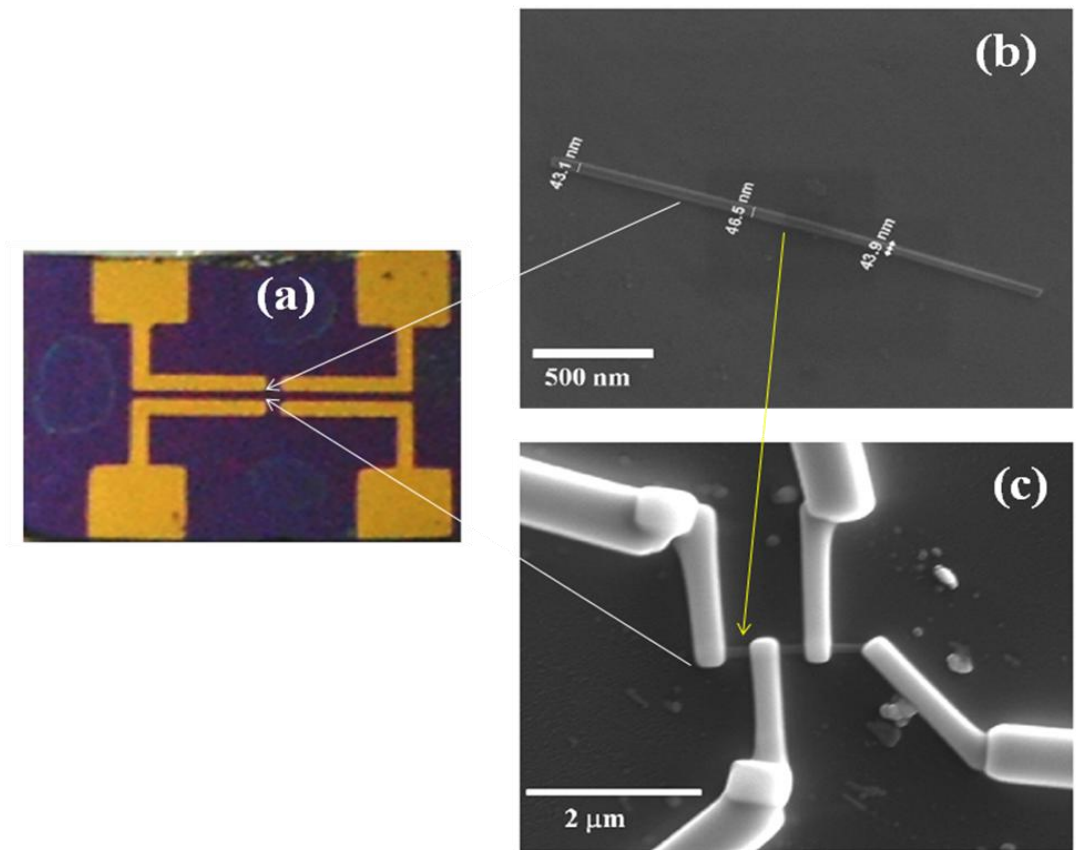

Fig. 6. (a) Cr/Au contact pads deposited on $SiO_2$ substrate by thermal evaporation using hard mask, (b) SEM image of single nanowire of diameter ~ 45nm, (c) SEM image of four-wire electrical contact made of Pt by using FIB. The white lined arrows in Fig. (a) indicate the area where the nanowires are dispersed and made contact with FIB. The yellow lined arrow indicates that nanowire is connected to four contact pads by Focused electron beam (FEB) and after that these are connected to gold contact pads by FIB.

*3.5.2. Transport measurement on single nanowire:*

The resistivity data are plotted in Fig. 7(a) as a function of temperature which were taken at a current of 1 µA.

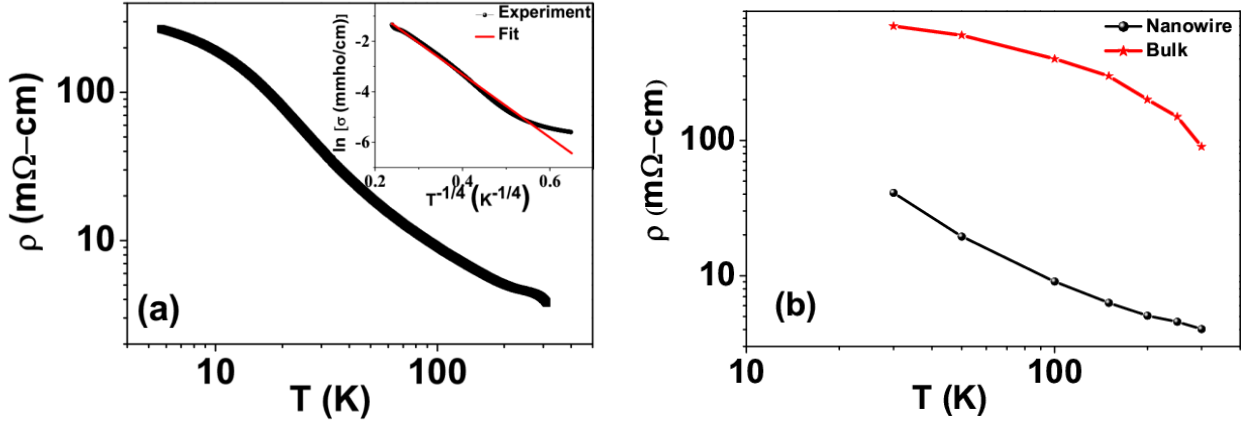

Fig. 7. (a) Temperature variation of resistivity of single nanowire of $La_{0.5}Sr_{0.5}MnO_3$ measured in the temperature range 5 K to 310 K. Inset curve (lnσ vs. $T^{-1/4}$) shows that conductivity data fitted with variable range hopping model for whole temperature ranges and (b) quantitative comparison of resistivity of LSMO nanowires and bulk [14].

We are able to make proper electrical contact and have measured the resistivity of single nanowire of LSMO (x = 0.5). Variation of resistivity as a function of temperature plotted in Fig. 7(a) shows insulating behaviour within the measured temperature range. The behaviour is similar to that seen in the bulk LSMO (x = 0.5) system [14]. The comparison of the resistivity of nanowire compared to the bulk ceramic sample is depicted quantitatively in Fig. 7(b). The bulk sample was prepared by ceramic method which shows quite higher resistivity than the nanowire. The ρ at 5 K is around $2.7 \times 10^{-3}$ Ω-m measured at 1 µA current and the resistivity at highest temperature (310 K) is $4 \times 10^{-5}$ Ω-m. We have fitted our conductivity data with variable range hopping model ($\sigma = \sigma_0 \exp(-(T_0/T))^{1/4}$) appropriate for whole temperature ranges (~ 10 K – 310 K) as shown in inset of Fig. 7(a) [17]. We have found the characteristic temperature ($T_0$) and density of states (DOS) at Fermi level ($N(E_F)$) from the fitted equation. From the slope of the fitted curve the estimated $T_0$ is ~ $2.4 \times 10^4$ K. $N(E_F)$ is related to $T_0$ by the relation $N(E_F) = \dfrac{24}{\pi a^3 k_B T_0}$, where a is localization length (~ 0.39 nm) which is around one unit cell. From measured $T_0$, we obtain $N(E_F) \sim 6.22 \times 10^{22}$ eV$^{-1}$cm$^{-3}$ or $1.55 \times 10^{43}$ J$^{-1}$mole$^{-1}$. From heat capacity data for a similar compound (LSMO x = 0.3) we obtain

N ($E_F$) ~ $0.96 \times 10^{43}$ J$^{-1}$mole$^{-1}$ [18]. The two DOS values are quite comparable and this validates the use of VRH model for the electrical conduction in the nanowire.

## 4. Conclusions:

We have demonstrated the fabrication of functional oxide nano wires of Sr doped lanthanum manganite system using autoclave and have used EELS as a tool to study the elemental composition of LSMO nanowires of diameter ~ 45 nm. From the EELS and magnetic measurements it has been observed that the nanowires retain the desired composition as well as basic ferromagnetic property, though $H_C$ increases even after the size reduction down to ~ 45 nm. Finally we are able to make four probe contacts to a single nanowire and perform the electrical transport measurements on a single nanowire down to 5 K. The resistivity from 310 K down to 10 K shows variable range hopping.

## 5. Acknowledgements:
B. Ghosh wants to thank DST, Govt. of India under UNANST Phase II for financial support. S. Datta would like to thank DST for financial support. The work has been done in the DST supported Unit for Nano science. S. Chandra and H. Srikanth would like to thank US Department of Energy (Grant no. DE-FG02-07ER46438), for the work at USF.